# Feeds as Query Result Serializations

Erik Wilde (School of Information, UC Berkeley)

**Abstract**

Many Web-based data sources and services are available as feeds, a model that provides consumers with a loosely coupled way of interacting with providers. The current feed model is limited in its capabilities, however. Though it is simple to implement and scales well, it cannot be transferred to a wider range of application scenarios. This paper conceptualizes feeds as a way to serialize query results, describes the current hardcoded query semantics of such a perspective, and surveys the ways in which extensions of this hardcoded model have been proposed or implemented. Our generalized view of feeds as query result serializations has implications for the applicability of feeds as a generic Web service for any collection that is providing access to individual information items. As one interesting and compelling class of applications, we describe a simple way in which a query-based approach to feeds can be used to support location-based services.

# Contents

# 1 Introduction

Feeds have become the most popular format for machine-readable data on the Web; they can be considered to be the most successful "Web service" available today. Feeds can be based on various data formats, either a considerable number of variations of the RSS format, or the more well-defined *Atom* [11] format. For the purpose of this paper, we do not look at the specifics of a particular feed format, but we do assume that feeds are using a well-defined format, and we will use Atom's terminology of a *feed* for the aggregate data structure, and of an *entry* as one item in that aggregate data structure. Based on the terminology of the *Atom Publishing Protocol (AtomPub)* [7], we use the term *collection* to refer to the data source underlying the feed, and a collection is a set of *members*. All of this means that a *feed* can be regarded as a standardized way to publish a set of *members* of a *collection* in the form of feed *entries*. Figure 1 is an illustration of this relationship between feeds, entries, collections, and members, which can be regarded as an example of the typical Web architectural pattern of resources (collections and members) and representations (feeds and entries).

In this paper, we investigate the applicability of feeds as a general format for query result serializations. We specifically address the missing parts of the current landscape of feed-related standards, and the question of how well feeds work as a general format for query result serializations. This investigation is a part of a more general question of how far the general concept of feeds can be taken, and Section 4 goes into more details about some of the scenarios in which feeds as a general format for query result serializations are a compelling design approach for for establishing a loosely coupled landscape of information and service providers.

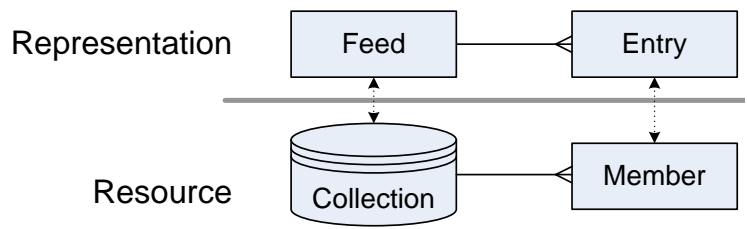

Figure 1: Relationship of Feeds, Entries, Collections, and Members

Generally speaking, we focus on feeds because in a view of Web engineering that is more centered on the Web than it is on back-end engineering [19], feeds stand out on today's Web as the most popular Web service in use. Feeds (in particular when AtomPub is taken into account as well) are a good example for how to use *Representational State Transfer (REST)* [6] as a design principle, and thus using feeds can be a promising approach when RESTful design is required.

# 2 Current Status

The current status of feed technologies on the Web can be regarded in terms of standards and standardization proposals (Section 2.1), and implementations (Section 2.2). Standards often lay the foundation and may explore interesting areas for generalized solutions, whereas implementations often are driven by the specific needs of an application or service provider. From the application area point of view, feed technologies started with RSS and then progressed to Atom when the RSS standardization situation became unsatisfactory for information providers interested in a stable, well-defined, extensible, and predictably evolving standard for feeds. The main application area for feeds from the end-user point of view are blogs and similar providers of time-ordered information streams (such as news agencies). Increasingly, feeds are used for any time-ordered stream of other types of content, for example photo management services such as *flickr*, *Picasa*, or *Panoramio* provide photo streams as feeds, allowing users to keep track of these streams with any service or tool supporting feeds.

In some of these scenarios, services are using their own conventions or extensions of feed technologies, but for all of these scenarios, the feed-related standards described in Section 2.1 are readily available.

## 2.1 Standards

The earliest instances of feed technologies were various RSS versions. RSS was riddled with a number of problems, caused by poor specifications (leaving crucial details to implementations), shifts in the basic syntax (some RSS versions are based on XML, others are based on RDF), disagreements in the community about the future path of the specification, and no clear stewardship for the specification itself.[1]

As a way out of this situation, *Atom* [11] (first named *Echo*) was created under the auspices of the IETF, and the main goals were a well-defined standard with unambiguous interpretation of the standard itself, and clear extension rules for how to extend the standard's basic model. In addition to the basic Atom model, there are standardized extensions for how to encode threaded discussions using *Atom Threading Extension* [16], how to embed licensing information in Atom feeds using the *Atom License Extension* [17], and how to publish various feed types for *Feed Paging and Archiving* [9]. These extensions cover scenarios that are recurring patterns for applications providing feeds, and allow standardized interactions around these recurring patterns.

Even though initially defined for RSS, *GeoRSS* [13] can also be used for Atom and is an extension for including geolocation information in feeds; the underlying geolocation model is the *Geography Markup Language (GML)* [12]. GeoRSS has two profiles, *GeoRSS Simple* and *GeoRSS GML*, with the former supporting point, line, polygon and box geometries, whereas the latter supports a wider array of features (such as alternative coordinate systems) and is formally defined as a profile of GML. Essentially, GeoRSS simply adds location information to the feed data model, but as we will see in Section 4, location as part of the data model is important in a wide variety of use cases for interacting with feeds.

The *Atom Publishing Protocol (AtomPub)* [7] is an extension of Atom which allows HTTP-based interactions with a provider of Atom feeds. While feeds are a read-only format, AtomPub extends the Atom feed model to cover all REST verbs, so that users can interact with an AtomPub-enabled collection by creating, updating, and deleting entries. While AtomPub is a useful extension of Atom's read-only model, in this paper we only consider queries and thus read-only accesses to Atom-enabled collections.

## 2.2 Implementations

For basic feed providers, Atom provides a sufficient foundation for publishing information, but as soon as information providers want to publish more sophisticated feeds for more specific access to the underlying collection, it gets more complicated. For example, one of the parts of Atom's data model are categories, and entries can be labeled with any number of categories. If information providers want to publish category-specific feeds, they can do so at different URIs, but then it is no longer clear that these feeds essentially are just different "views" of the same underlying feed or collection. Some blog services, for example, provide users with the ability to publish category-specific feeds, and blog posts are published in the feeds for the categories that were used to tag them, as well as in the general feed.

This situation becomes even more difficult when users should be allowed to subscribe to feeds with more than one category assigned to it; in this case, the feed provider has to invent some convention for expressing the categories contained in the feed.[2] The technology Q&A site `stackoverflow.com`, for example, has introduced a custom syntax for this, allowing feed URIs to carry query information and supporting *and*, *or*, and *not* operators, using a simple syntax:[3]

---

[1] http://diveintomark.org/archives/2004/02/04/incompatible-rss claims that there are at least 9 incompatible versions of RSS, based on XML vs. RDF, different timestamp formats, different metadata models, and different escaping rules for embedded content.

- *java* and *jsp*: `http://stackoverflow.com/feeds/tag/java jsp`
- *java* and not *jsp*: `http://stackoverflow.com/feeds/tag/java -jsp`
- *java* or *jsp*: `http://stackoverflow.com/feeds/tag/java or jsp`

While this provides useful functionality, this solution has two drawbacks: it may compromise a service's scalability (more on this topic in Section 3.1), and it is specific for this one service, so that clients cannot easily support this feature programmatically across different providers.

Google's *GData* protocol combines Atom and AtomPub, and also covers query features for feeds. It is intended as a general API for accessing many of the data sources and services provided by Google. GData uses a different syntax than the one presented above, but very similar functionality. The GData syntax for the above feed URIs would look like this:

- *java* and *jsp*: `http://stackoverflow.com/feeds/-/java/jsp`
- *java* and not *jsp*: `http://stackoverflow.com/feeds/-/java/-jsp`
- *java* or *jsp*: `http://stackoverflow.com/feeds/-/java|jsp`

In addition to categories, GData supports full-text queries, author searches, range queries on the *updated* and *published* timestamps, and methods for specifying the results as number of entries, the start index of the returned results, and the returned data format. The GData query model is more a filter than it is a full query: all query predicates are scoped to a single entry, and thus the complete query acts as a filter for each entry individually. One apparent limitation of GData is that it allows queries on nothing else but the data exposed in a feed (more on that distinction in Section 5, which discusses the difference between *querying feeds* and *querying collections*).

An alternative approach is the *Feed Item Query Language (FIQL)* [10], which also specifies a query syntax for feeds, but defines a more general set of operators, datatypes, and values which can be used for queries. Most importantly, FIQL defines a way how a feed provider can advertise its query capabilities in the feed itself using a *query interface*, which means that a feed can advertise the query capabilities of the underlying collection. While FIQL is an interesting approach and heads in an interesting direction, the initial and expired draft was never updated, and no known implementations exist.

Two other implementations supporting query-oriented interactions with feeds are *Yahoo Pipes* and the *Yahoo Query Language (YQL)*. Pipes is a visually oriented tool for how users can connect various feed-processing modules, and can aggregate and filter feed data (or other data providers which are turned into feeds) to generate one result feed. The fundamental limitation of Pipes is that it only works on the feed level, not on the collections, because the platform is designed as an intermediary between feed providers, and the consumer of the feed generated by the platform. YQL also supports various data input (including feeds), but the output is YQL-specific XML or JSON. So while YQL targets the same goal as GData, a uniform access to various data sources provided by the respective companies, it is not using feeds as its fundamental data model.

---

[2]Publishing all permutations of categories as individual feeds becomes impractical for more than a very small number of categories.

[3]All examples of URIs shown here are not properly URL-encoded. This makes it easier to recognize the characters used for the URI syntax.

Even though it is not strictly feed-based, *OpenSearch* is another widely recognized approach to consider. It was developed by *A9*, a subsidiary of *Amazon*, and mostly is a standardized API for search engines. The interesting fact is that OpenSearch uses feeds with extensions as the result format. As the query format, OpenSearch defines XML formats (one for describing the interface, and one for the query itself), as well as URI templates. The main conceptual limitation of OpenSearch is that it is focused on full-text search.

## 3 Implementation Issues

One of the most important reasons for the success of feeds is that they are an excellent example of *loose coupling* [14]. They provide a resource-oriented and easily usable way of accessing a great variety of service providers, and allow both providers and consumers to use simple implementations. Even though colloquially speaking people often talk about "subscribing to a feed", there actually is no such thing as a feed subscription; servers provide feeds, and clients read feeds. This approach provides looser coupling than publish/subscribe approaches [5] (which forces publishers to keep track of all subscribers and deal with stale subscriptions), and allows servers and clients to benefit from standard Web mechanisms for load balancing, such as HTTP intermediaries (which might or might not act as caches).

### 3.1 Server-Side

The most important benefit from the server-side point of view is the almost perfect *scalability* of feeds. Since feeds are often provided as static Web resources, servers can refresh these resources as required, and they can be served as static documents in between updates. This is possible because all clients get the same feed when they request it; there is no such thing as a client-specific feed. This "one size fits all" approach allows servers to scale to millions of feed consumers with little effort, because feed requests by clients do not cause any back-end query. Feed providers can use standard Web methods and appliances to improve performance, should scalability become an issue.

Purely category-specific feeds (one feed per category) could be regarded as a "few sizes fit even more" approach, which could still be easily implemented with static feeds, which are only refreshed when new entries are published. As long as individual client requests are not mapped to individual back-end queries, feeds scale well. On the other end of the extreme are individual feed queries, which are the equivalent of the observation that "personalized sizes are expensive to provide", at least as long as the personalization process is not optimized. If individual feed queries are naïvely mapped to back-end queries, scalability suffers (Section 5.1 goes into more detail about this question).

However, even with naïve implementations it is possible to imagine scalable scenarios, where free access to feed-based data sources is provided with the current non-queryable feed model; and as a value-added service it is also possible to get query access to the feed for more advanced applications. However, this query-enabled access would be controlled through mechanisms such as HTTP authentication or API keys for enforcing access limits and/or paid subscriptions; thereby giving the feed provider the ability to better control feed access through a two-tier access scheme (read-only access and query access).

### 3.2 Client-Side

While the main advantage of the simplicity of the current feed model lies with servers and their ability to scale easily, clients also gain some benefits from feed processing. While there is no guarantee that the entries in feed are ordered by time (which would allow clients to do more optimized processing), Atom does guarantee that feed entries have unique and stable identifiers. By keeping track of these identifiers, clients can synchronize the entries in a feed they read with previously read entries, and update their local storage.

If a feed supports *Feed Paging and Archiving* [9], clients can interact with a feed in a manner guaranteeing that they will not loose any entries. This Atom extension supports *archived feeds*, which essentially are stable feeds which are supposed to contain all the entries of a logical feed. Typically, clients reread feeds as required, and they do not necessarily care about the fact whether they have missed some entries because the feed was updated faster than they were rereading it. However, if a feed supports archiving, clients interested in the complete history of a feed can always retrieve the complete history of a feed by following links to archived feeds.

Client-side processing and optimizations often look at the current model of feeds, which is based on the implicit assumption that updates will appear in feed, and that by rereading a feed at an appropriate interval, clients can get a complete picture of the updated entries for the underlying collection. In Section 4, we look at scenarios where the collection itself, or the interactions with a collection, do not map well to this model of feeds, and where an explicit query mechanism for feeds would significantly improve the applicability of feeds.

It should be kept in mind, though, that *Feed Paging and Archiving* already is a first step in that direction: the ability to have paged and archived feeds already establishes a richer set of interactions between feed providers and consumers, even though in this case this is not implemented in a query language, but by using special link relationships. It should be noted that many feed providers supporting feed paging already do so by linking these link relationships to "pseudo-query" URIs such as "`...feed.xml?page=2`" when linking to the second page of a paged feed.

## 4 Feed Query Scenarios

The previous sections have pointed out how feeds today are mostly used in a "one size fits all" approach, and how this pattern so far has been only partly broken by proprietary implementations such as GData and YQL, or by specialized extensions such as *Feed Paging and Archiving*. The current main scenario is based on feed access regardless of the specific needs of the feed consumer. *Location-Based Services (LBS)* [15] are a popular example where access to an information source or service is location-dependent. In this case, the location of a feed consumer should result in different responses of the service provider, depending on the resource.

As one example, it is possible to envision a general feed reader (similar to, for example, the popular Google Reader) aggregating several feeds, but these feeds are location-enabled and support location-based access. Such a "geofeed reader" would probably use a map-based interface or at least some map view of the retrieved information.[4] In such an interface, the map view could be regarded as an "implicit query", so that user interactions with the map view would cause the feeds to be queried for the new view. Such an interface would be particularly interesting for mobile devices, which in many cases are used in LBS scenarios, and which also can benefit from the significant bandwidth optimization of only requesting those entries which are relevant to the current location or map view.

While the LBS scenario in combination with query-enabled geofeeds alone is a rather compelling use case, more complex models of user interests and situations than just the location [8] can also be employed. Mapping such a more complex situational model to feeds and feed queries (if they have various query capabilities) becomes a flexible and powerful way of service personalization. Feeds could be pulled from a large variety of providers, more traditional feed publishers such as news agencies, but also information from sources such as traffic sensors (when the situation is recognized as requiring traffic information). Generally, in such a scenario the situational context can be seen as a query into an aggregated set of feeds, and the expected response is the aggregated set of results from all of these services, providing the most useful information for the current situation. The hard challenges in this scenario are figuring out how to know which feeds to aggregate, and how to map the situational context to the query capabilities of those feeds; once these challenges have been met, using feeds as the serialization format establishes a common and loosely coupled foundation for the feed providers and consumers to cooperate.

---

[4]Chen et al. [3] describe such a system with a specialized user interface, which is built on a proprietary platform for providing spatial and temporal query services.

Feeds are often regarded as a serialization of inherently time-ordered data, but there is nothing in the Atom specification that prevents a feed from being used as a way to serialize data that has other primary keys than a timestamp (the only thing required by the specification is that each entry must have an *updated* timestamp, indicating when an entry was last updated). In addition, the specification assigns no significance to the order of entries within a feed. Consequently, a service responding to a feed query with a result feed can choose to sort the results in any way, and most importantly, if the query language also supported features such as *sort-by* and *group-by* constructs, the results could be ordered according to these parameters.

## 5　Query Models

Based on the scenarios presented in the previous section, this section looks at considerations for the specific query data model and language. Sections 5.1 to 5.3 look at the question of the extent of such a query language, i.e. at the question of what should be queryable. Sections 5.4 and 5.5 then look at the issue of the query language itself and how it could support queries joining entries from one feed, or even across aggregated feeds.

Some of the implementations discussed in Section 2.2 have query data models that are specific for specific data sources. GData and YQL have been designed as unified interfaces for various services provided by the respective companies, and thus have well-defined ways of how information specific to these services can be retrieved using these languages. However, this means that these services are essentially hard-coded into the languages' specifications, defining how specific services can be used with these languages. While this makes sense as an approach for integrating a rather small and centrally controlled set of services, it is not an approach that readily translates into an open scenario.

Furthermore, none of the implementation described in Section 2.2 cover geolocation as part of their data model, and we believe that geolocation is a concept important enough that it should be part of the standard interactions with a feed-based collection. Sections 5.1 and 5.2 explore the basic difference between interactions only using data contained in the feed, or interactions which also can use additional data contained in the collection.

The question of the best data model and query language for feeds has quite a bit of similarity to the problem of *federated databases*, but there are some relevant differences. One point is that it can be assumed that the collections underlying feeds are using very different data models, and these data models are also based on different metamodels. On the other hand, since these collections are published as feeds, there is some mapping from their data models to the feed data model. This means that on the one hand the feed data model is useful because it provides a universal metamodel across feeds, but this also means that for many collections, only parts of the collection's model will be mapped to feeds, while other are "invisible" on the feed level.[5]

This unified view of collections introduces an interesting opportunity for large-scale information mining. Individual collections can expose services which are specifically designed for these collections, and cannot be used outside of their scope. On the other hand, feeds can be aggregated and then they become "view" of these individual collections, represented in a unified way. This aggregated information can then be used to perform cross-collection queries which would have been impossible when dealing only with individual collections. Section 5.5 goes into more details about the query features required for this scenario, but it is worth pointing out that the two levels of collections and feeds, of implementation-specific and standardized data formats, enable new forms of data filtering and aggregation that would be impossible with a less loosely coupled approach.

---

[5]Such additional information in the collection can be entirely collection-specific and not be represented in the feed at all, or it can belong to a scenario that is popular and has its metamodel exposed using Atom extensions. One popular example for this are *podcasts*, which are regular feeds, but use a number of extension elements to represent the additional information that is required for the specifics of making episodes of audiovisual content available.

## 5.1 Feed Queries

The most limited data model for feed queries is one that only takes the standardized fields of feeds into account. According to the Atom specification, a feed can have 12 possible element types as children that will describe the feed itself, and each entry in a feed can also have 11 possible element types (almost the same ones as the feed) as children that will describe the entry. The entry element either contains or links to the actual *content*, which can be *text*, *HTML*, *XHTML*, or some MIME media type. In Atom's metadata elements, various data types are used, most notably *text*, *timestamps*, *URIs* (Atom actually uses IRIs), and *email addresses*.

The elements and datatypes defined by Atom itself are the most restricted subset that could be considered for querying feeds (if we do not consider pure full-text search approaches such as OpenSearch). Based on the datatypes, query operators for these elements could include complete and partial matches for text, URIs, and emails, and range queries for timestamps. For text, more advanced query methods could be full-text search methods such as thesaurus-based search, stemming, case-control, and support for diacritics.

Such a query language operating on the feed data alone could be implemented by mapping it to *XQuery* [2] (possibly extended by XQuery's full text extension [1], if full-text search is considered as well). Depending on the underlying collection, though, for the feed provider it would be the more efficient option to map a feed query to the query language of the back-end, so that the query could be optimized as much as possible.

One important issue to point out is that the query model presented here, as well as the one presented in the following section, is only operating on the feed level. This means that an intermediary with no knowledge of the back-end of the feed provider is able to execute the same queries; limited of course to those entries that have been cached by that intermediary. While those intermediaries can be any HTTP-level intermediaries, there are also scenarios where a feed provider specifically decides to publish their feed through an intermediary, usually for load balancing and/or analytics. One such example is *FeedBurner*, a company that is now part of Google. They act as intermediaries for feed providers, and cache feeds (though not necessarily the complete history of a feed) and thereby greatly reduce the load for feed providers. If such a intermediary does archive feed entries, however, it can also process feed queries based on this archive.

## 5.2 Extended Feed Queries

While Atom itself uses the rather small set of element types mentioned in the previous section, it also has a well-defined extension model, and extension elements can appear within feeds as well as within entries. One such extension is GeoRSS markup, which encodes geospatial information in a feed. In a strictly feed-based model as the one described in Section 5.1, this information cannot be queried, because it is not part of Atom itself. However, there are useful extensions to Atom specifying additional elements for feeds and entries, GeoRSS is one of them, the *Atom License Extension* [17] is another one.

In both of these examples, the extensions specify a way in which additional information can be embedded in a feed, essentially extending the information that may appear in the feed, as well as the data model. In case of GeoRSS, this is especially important because it adds the complexity of the GML model to feeds, whereas licensing information is made available through a new link relation type, and thus is reduced to a URI/IRI in terms of its contributions to the information in the feed itself.

These examples thus show the breadth of problems that may occur when considering the ability to add query capabilities for feed extensions. For license information, this essentially means to query for standard *link* elements, but only those using a standardized relationship. Thus, if the query language supports queries on links that take into account the relationship, the relationship's namespace, and the link's target, querying for specifically licensed entries can be done.

On the other hand, for GeoRSS, the feed extension introduces new elements and new data types, and for querying to be able to work on this as well, the query method must be more radically extended.

Since we assume that query capabilities will always be based on the collection being published in a feed, it would make sense to define extended query capabilities as separate modules, so that an extension such as GeoRSS would have an accompanying specification defining the query capabilities for this extension. For this to work, the query language must have a well-defined and modular extension system.

While GeoRSS is used by a large set of feed providers, such a model would also allow service-specific extensions to fit into the framework of the query language, so that service providers such as Google and Yahoo could define and publish their own modules. Reusable modules would be the preferred way, though, because then feed consumers can interact with any service providing feeds and query capabilities with that functionality.

One interesting approach taken by FIQL is to allow feed queries to use *XML Path Language (XPath)* [4] expressions to identify values that can used in queries. While this makes it possible to select markup constructs such as feed extensions, and thus introduces some extensibility into the query language, in case of GeoRSS this would be of little use, because selecting GeoRSS elements will return GML, and thus this datatype (which allows rather complex geometric shapes to be used) must be known to be able to process these values.

More generally speaking, while it is possible to create an extensible feed query language that does support basic feed elements and datatypes as described in Section 5.2, and can be extended in a modular way, there still is no way how feed providers can expose services beyond the information contained in the feed. This is good from the standpoint of intermediaries, because this means that whatever feed consumers can get from a feed provider, they can also get from an intermediary; but it ignores the fact that the collection itself may have more capabilities than those represented in its feed serialization.

## 5.3   Collection Queries

The feed queries described in the previous sections are limited to the data that is available in a feed, they cover standard elements of Atom (Section 5.1) and extensions (Section 5.2). The next logical step is to expose the query capabilities of the collection, which may go beyond what is represented in the feed.

As an example, a collection of images could be exposed through a queryable feed, and users of that collection could query for images using standard Atom fields (such as a title or a timestamp), or they could query for images using information that is available in the feed as extensions (such as the location where a picture was taken, or the license under which it has been published). However, more advanced services such as content-oriented image search could also be provided, giving users the opportunity to search for images by textual descriptions of what they are looking for, or maybe by providing example images.

This query capability of the service is not at all exposed in the feed itself, but it may be an important service provided by the collection. Thus, including query capabilities which not only extend Atom's data model and datatypes, but also have no representation in the feed itself once the query has been processed, could greatly expand the possible application areas of feed-based querying.

It should be kept in mind, though, that such a collection-based query is something that cannot be supported by intermediaries, because these have only access to the feed data, but not to the collection. Thus, if a service provider is interested in supporting intermediaries on the feed level (for example for load balancing), then collection queries cannot be used. If collection queries are used, queries must be run on the collection in the back-end, and possible load balancing then has to be done on that level as well.

## 5.4 Cross-Entry Queries

The query models described in the previous sections are of increasing complexity regarding the ability to query data beyond the feed data model itself, but they do not necessarily extend the query model beyond simple filtering capabilities. The next possible level of complexity is that of queries relating entries to each others; in relational databases, this operation is often referred to as a *join*, and since a feed is just one data structure, more specifically it could be regarded as a *self-join*.

Since collections accessible through feeds often do have timestamps associated with them, a special case of self-join would be *temporal queries*, which often are queries such as "find all entries where there were more than three entries published over the course of one hour".

Another interesting special case of self-joins are *spatial queries*, which essentially can exhibit the same nature than temporal queries, but extend time's one-dimensional model to the 2D or 3D space of the underlying spatial model. Spatial queries often require the same self-join logic as temporal queries, asking questions like "find all entries in places where there are more than a certain number of entries in a given area".

Combining temporal and spatial queries leads to the field of *spatio-temporal databases*, which is a wide and well-established field of research and productive systems. More specifically, sensor networks add an additional complication to this field by often using models of *streaming data*, so that spatio-temporal queries must be processed in real-time and over potentially a large number of incoming datastreams. Sensor networks typically have a large volume of incoming data (the "collection", which in this case is a continually updated set of datastreams), and ideally should only produce a small volume of outgoing data (the feed of relevant events). In such a network, queries often are used for alerting systems, so that for example a sustained condition over a certain period of time can trigger an alert.

## 5.5 Cross-Feed Queries

Using feed query capabilities for querying a given sensor network alone already can provide substantial benefits by providing a uniform interface for different underlying implementations of sensor networks [18]. However, more interestingly, if the results produced by different sensor networks can be easily aggregated and processed, even more opportunities for using sensor networks become visible.

As mentioned in Section 5, one very important use case for cross-entry queries are *aggregated feeds*. If feeds are regarded as query result serializations of individual services, these results can be aggregated and then queried again. This kind of query can be regarded as a *join* across multiple feeds, allowing it for example to use the aggregation of two geofeeds, and query this aggregation for spatial co-occurrences of entries across the feeds, thereby joining the feeds in a spatial manner.

This last scenario for cross-entry queries would make it necessary to augment a feed with the *origin* for each individual entry, so that aggregated feeds still contain the information in which feed an individual entry originally occurred. We call such an aggregated feed a *feedset*, and while a feedset is just a feed, the interesting fact about is that its entries correspond to members from more than one collection, turning it into a representation of resources across various collections. Figure 2 shows an example of a feedset, and it is important to note that the various collections represented by a feedset are not necessarily related by anything more than the fact that both of them are accessible through feeds.[6]

When considering *cross-entry queries* as discussed in Section 5.4, and *cross-feed queries* as a even more powerful concept for feed queries, it is important to consider the implementation options for such a language.

---

[6]This means that it is possible to implement cross-feed queries by handling the individual query parts upstream, i.e. closer to the original data source (if the aggregated feeds are query-enabled), but that it will not be possible to do this if the query has join semantics across feeds. If that is the case, this creates the same situation that led to the invention of *data warehouses*, where such a cross-feed query could only be efficiently processed in the presence of a "feed warehouse" (effectively the intermediaries mentioned in Section 3).

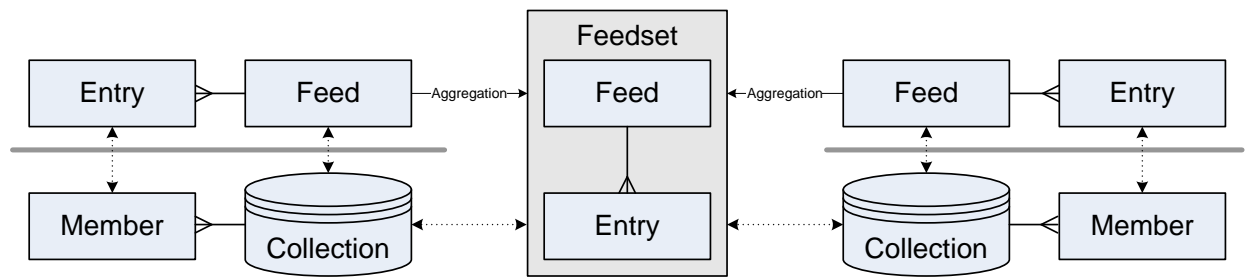

Figure 2: Feedsets are Feeds which are aggregating Feeds into new Feeds

Considering *feed providers*, *intermediaries*, and *feed consumers* in such a scenario, the following observations can be made:

- Pure filtering capabilities as well as cross-entry queries can be performed by the feed provider. They could also be performed by intermediaries (as long as queries are not collection-based, in which case *only* the feed provider can process them), but in terms of optimization, pushing them as far upstream as possible is preferable.

- Cross-feed queries, on the other hand, cannot be performed by feed providers, because the very nature of these queries requires an aggregation of feeds before the queries can be performed. This means that the model of intermediaries is a very useful one for these queries, because queries can then be pushed as far upstream as possible (i.e., to the intermediary performing the aggregation).

This means that for query optimization and when looking at a scenario where a feed consumer is aggregating various feeds, query processing should be able to be done in a distributed fashion, so that filtering of feed entries can be done as far upstream as possible, reducing the bandwidth and processing load required further downstream.

## 6 Discovery

It is also important to consider how a feed's query features can be discovered. For today's feeds, no discovery is required, because a feed is not query-enabled. However, *Feed Paging and Archiving* (as described in Section 3.2) already does introduces minimal "query features" for a feed, and does so by adding special links to a feed. This means that discovery of that feature is entirely feed-based; a client can simply retrieve a feed and will understand whether that feeds support paging and/or archiving or not.

The same principle should be used for the more complex query features discussed in Section 5. FIQL and OpenSearch both define ways in which a client can discover the query features of a service provider. The mobile map-based geofeed client described in Section 4 should be able to retrieve a feed, figure out the feed's query capabilities, and query that feed according to the feed's capabilities. The query can be driven by the current map view, some internal situational model, or an explicit user interface that is generated based on the feed's query capabilities. The important benefit of this design is that no prior configuration is required, and providers and consumers can interact spontaneously.

## 7 Conclusions

Feeds are the most prolific Web service available on the Web today. Based on a simple and loosely coupled model, they allow feed consumers to interact with a variety of service providers. By looking at the current feed model as a simple version of how feed consumers get representations of queries run on the feed provider side, it is possible to envision extensions of that simple model. Based on the current standardization and implementation landscape, and taking implementation issues for both providers and consumers into account, an extension of the current model of "hardcoded" queries to a more flexible and powerful model can be envisioned.

Generally speaking, the view of feeds as query result serializations, and a possible extension of the feed format to provide query support, is an attempt to allow declarative ways of how feed consumers and providers can interact. The detailed implementation issues for query-enabled feed services are outside the scope of this paper, but many research results from focused research areas such as federated databases, distributed query processing, sensor networks, and spatio-temporal databases could be leveraged to address many of the challenging research questions.

It should be noted that while in this paper we have not looked at issues of authentication, access control, and security, many of the standard issues in this areas could be readily addressed by using standard Web technologies.